\RequirePackage{fix-cm}
\documentclass[twocolumn,epjc3]{svjour3}  
\smartqed  
\RequirePackage{graphicx}
\RequirePackage[colorlinks,citecolor=blue,urlcolor=blue,linkcolor=blue]{hyperref}
\begin{document}

\title{Low Gain Avalanche Detectors for the HADES reaction time (T$_0$) detector upgrade
}
\subtitle{\small{Published in Eur. Phys. J. A (2020) 56:183 \\ \url{https://doi.org/10.1140/epja/s10050-020-00186-w }}}

\author{
J.~Pietraszko$^{1,}$\thanksref{e1}\and
T.~Galatyuk$^{1,2}$\and
V.~Kedych$^{2}$\and
M.~Kis$^{1}$\and
W.~Koenig$^{1}$\and
M.~Koziel$^{3}$\and
W.~Kr\"uger$^{2}$\and
R.~Lalik$^{4}$\and
S.~Linev$^{1}$\and
J.~Michel$^{3}$\and
S.~Moneta$^{5}$\and
A.~Rost$^{2}$\and
A.~Schemm$^{6}$\and
C.J.~Schmidt$^{1}$\and
K.~Sumara$^{4}$\and
M.~Tr\"ager$^{1}$\and
M.~Traxler$^{1}$\and
Ch.~Wendisch$^{1}$
}
\institute{$ $ GSI Helmholtzzentrum f{\"u}r Schwerionenforschung GmbH, 64291 Darmstadt, Germany\\
 $ ^{2}$ Technische Universit{\"a}t Darmstadt, 64289 Darmstadt, Germany \\
 $ ^{3}$ Institut f{\"u}r Kernphysik, Goethe-Universit{\"a}t, 60438 Frankfurt, Germany \\
 $ ^{4}$ Smoluchowski Institute of Physics, Jagiellonian University of Cracow, 30-059 Krak{\'o}w, Poland \\
 $ ^{5}$ Universit{\`a} di Pisa, 56126 Pisa, Italy \\
 $ ^{6}$ IMT Atlantique, Campus de Nantes, 44307 Nantes, France
}
%
\thankstext{e1}{e-mail: j.pietraszko@gsi.de}



\date{}

\maketitle

\begin{abstract}
Low Gain Avalanche Detector (LGAD) technology has been used to design and construct prototypes of time-zero detector for experiments utilizing proton and pion beams with High Acceptance Di-Electron Spectrometer (HADES) at GSI Darmstadt, Germany.
LGAD properties have been studied with proton beams at the COoler SYnchrotron (COSY) facility in J\"ulich, Germany.
We have demonstrated that systems based on a prototype LGAD operated at room temperature and equipped with leading-edge discriminators reach a time precision below 50~ps.
The application in the HADES, experimental conditions, as well as the test results obtained with proton beams are presented.


\end{abstract}
\section{Introduction}
\label{intro}
The HADES collaboration~\cite{Ref_Hades} is developing a new time-zero (T$_0$) and beam tracking system for the upcoming experiments that will use proton and pion beams~\cite{Ref_Hades_pion_facility}. 
This system is meant to replace the currently utilized single-crystal chemical vapor deposition (scCVD) diamond based detector that, although successfully used for these purposes~\cite{Ref_Hades_di_mips,Ref_Hades_dia}, possess a number of limitations, 
like $e.g.$ small sample sizes, with typical dimensions not larger than 5~mm $\times$ 5~mm, that do not allow to build large area detectors.
The newly available sensors based on the Low Gain Avalanche Detector technology (LGAD)~\cite{Ref_Design_ufsd,Ref_Sarodz_LGAD,Ref_First_FBK}, aka Ultra Fast Silicon Detectors (UFSD), provide excellent position measurement capabilities and additionally a fast signal response with a precision better than 100~ps~\cite{Ref_Beam_ufsd_16ps}.
These properties, combined with high radiation hardness~\cite{Ref_RadHard_ufsd} and low production costs, are very attractive for tracking and timing applications.
A demonstration system has been realized as a beam telescope consisting of the two LGAD strip sensors, which was exposed to a proton beam at the COSY Synchrotron at J{\"u}lich, Germany. 
This paper describes the experimental setup, the data readout system, and details of the analysis.
Particular emphasis is put on the results obtained with these prototype sensors, further steps towards the final system for production experiments with HADES are summarized in the outlook.
\section{T$_0$ detector requirements for Minimum Ionizing Particle (MIP) beams}
\label{sec_det_req}
The HADES fixed target experiment is located at the SIS18 (heavy-ion synchrotron with rigidity 18 Tm) accelerator in Darmstadt, Germany, and investigates microscopic properties of resonance matter formed in heavy-ion collision in the 1-2$A$ GeV energy regime, as well as exclusive channels in proton and pion beam induced reactions, both for hadronic and semi-leptonic final states.
For the physics program of HADES, it is necessary to determine the T$_0$ of the reaction with a precision better than 60~ps ($\sigma_{\textrm{T}_0}$) and monitor the properties of the beam, in particular its position, width, and time structure.
The measured T$_0$ defines the start time of the time-of-flight (ToF) measurement needed for particle identification.
To ensure full control of the beam position on the HADES target, it is necessary to track each individual beam particle in front of and behind the target, and determine the beam parameters close to the HADES focal point. 
In addition, it should be noted that, for fixed-target experiments with beam intensity of up to 10$^8$ particles/s, not only information about the position of the particle but also information about the arrival time is of great importance.
Performing spatial track reconstruction using, in addition, the time information (4D tracking) will significantly reduce the number of chance coincidences, $e.g.$ wrongly correlated hits.
A 4D particle tracking is proposed in many experiments, $e.g.$ at the high-luminosity LHC~\cite{Ref_Ufsd_4d_track}. 

For the upcoming HADES experiments with pion and proton beams, the T$_0$ detector has to fulfill the following requirements:
\begin{itemize}
\item Good timing precision with $\sigma_{\textrm{T}_0}<$60~ps for particle identification via time-of-flight.
\item Operation for particle fluxes of $J>10^7$~p/(cm$^2$s).
\item Detection efficiency for MIPs close to 100\%.
\item Low material budget, below 0.5~mm Si equivalent.
\item Position determination capabilities of $\delta$x$<$0.5~mm.
\item Vacuum operation capability.
\item Active area of up to 8~cm$^2$.
\end{itemize}

Detectors based on LGAD sensors are great candidates to satisfy these demands.
To investigate the capabilities of such detectors, several small demonstration systems realized as a beam telescope have been designed, constructed, and tested using proton beams of 1.92~GeV kinetic energy. 

\section{Experimental Setup}
\label{sec_setup}
\begin{figure}[h]
\begin{center}
  \includegraphics[width=0.45\textwidth]{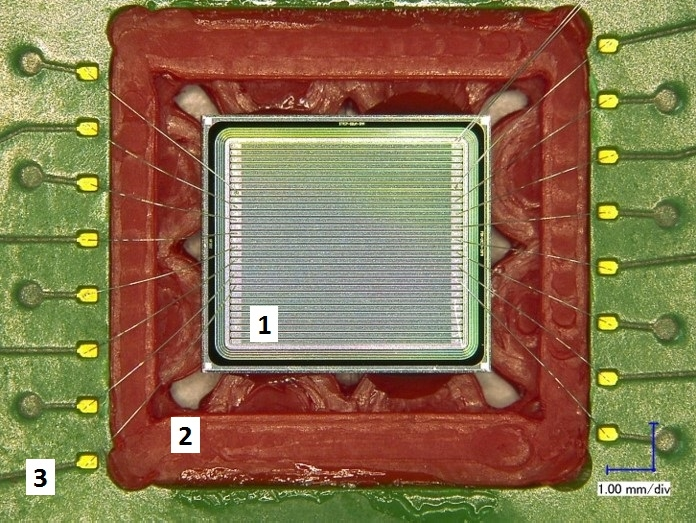}
\end{center}
\caption{Photograph of the LGAD sensor~(1) with size of 5.0~mm x 4.3~mm mounted on a PCB plate~(3) with the help of an adapter~(2) to ensure the correct sensor positioning. 16 of 30 readout strips were bonded to the PCB traces.}
\label{fig_ufsd_on_pcb}       
\end{figure}

\begin{figure}[h]
\begin{center}
  \includegraphics[width=0.45\textwidth]{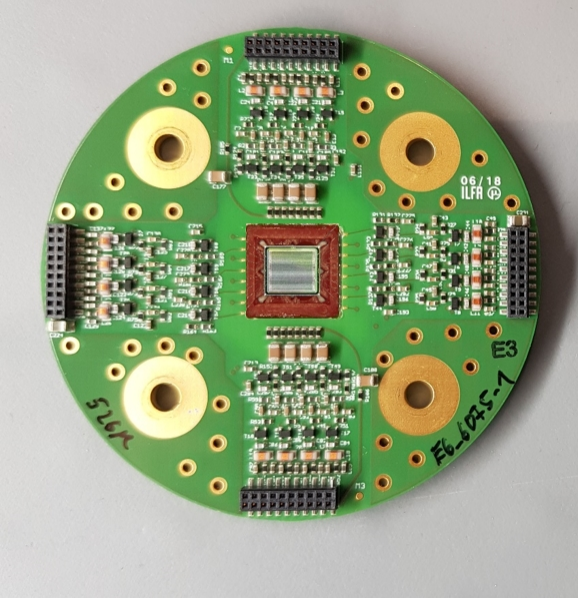}
\end{center}
\caption{Photograph of the LGAD sensor mounted on a PCB plate with 5~cm diameter designed for double-sided strip sensors. The PCB shown in the photo contains 32 channels organized into groups of 8 oriented in 4 different directions. For single-sided LGAD strip sensors only the 16 front-side channels from 32 available on a singe PCB were used.}
\label{fig_ufsd_on_pcb_large}       
\end{figure}

Two different T$_0$ detector types were built.
One based on the LGAD sensors with Boron type gain layer implant (called W3) and one where the gain implant was based on gallium with low amount of carbon (called W15)~\cite{Ref_First_FBK}.
Both types of the sensors featured a form factor of 5.0$\times$ 4.3~mm$^2$ and were manufactured at Fondazione Bruno Kassler (FBK).
The LGAD sensors are single-sided, multi-strip devices with an active thickness of 50~$\mu$m, a pitch of 146~$\mu$m and a strip-to-strip distance of 20~$\mu$m.
The segmentation of the LGAD sensor is shown in Fig.~\ref{fig_ufsd_on_pcb}. 
As investigated in details using the X-ray focused beam technique~\cite{Ref_Xray_test} these sensors have a low electric-field region between neighboring gain implants of about 90~$\mu$m width where the measured gain is at least 50$\%$ lower than the maximum.
The low electric-field region (low gain region) can be slightly reduced by increasing the bias voltage.
By extrapolating the measurement shown in \cite{Ref_Xray_test}, Figure 9, the fill factor of these LGAD sensors can reach about 55-60\% at a gain of 20 and a bias voltage of about 300V.
Such a low fill factor prevents the direct use of these prototype sensors in the experiments, nevertheless it allows accurate testing of their time properties and rate capabilities.
The prototype T$_0$ detectors consisted of the LGAD sensors glued to Printed Circuit Boards (PCBs) which served as a holder for the sensor, contained the front-end electronics, and provided electrical connections to the readout system.
The readout pads of the LGAD sensors were wire bonded with dedicated traces onto the supporting PCB (only 16 strips located in the middle of each sensor were bonded for these studies).
The signals were next processed by the two stage amplifiers (more details in Fig.~\ref{fig_ufsd_on_pcb_large} and reference \cite{Ref_Hades_di_mips}) followed by  the discriminator boards.
\begin{figure}[h]
\begin{center}
  \includegraphics[width=0.45\textwidth]{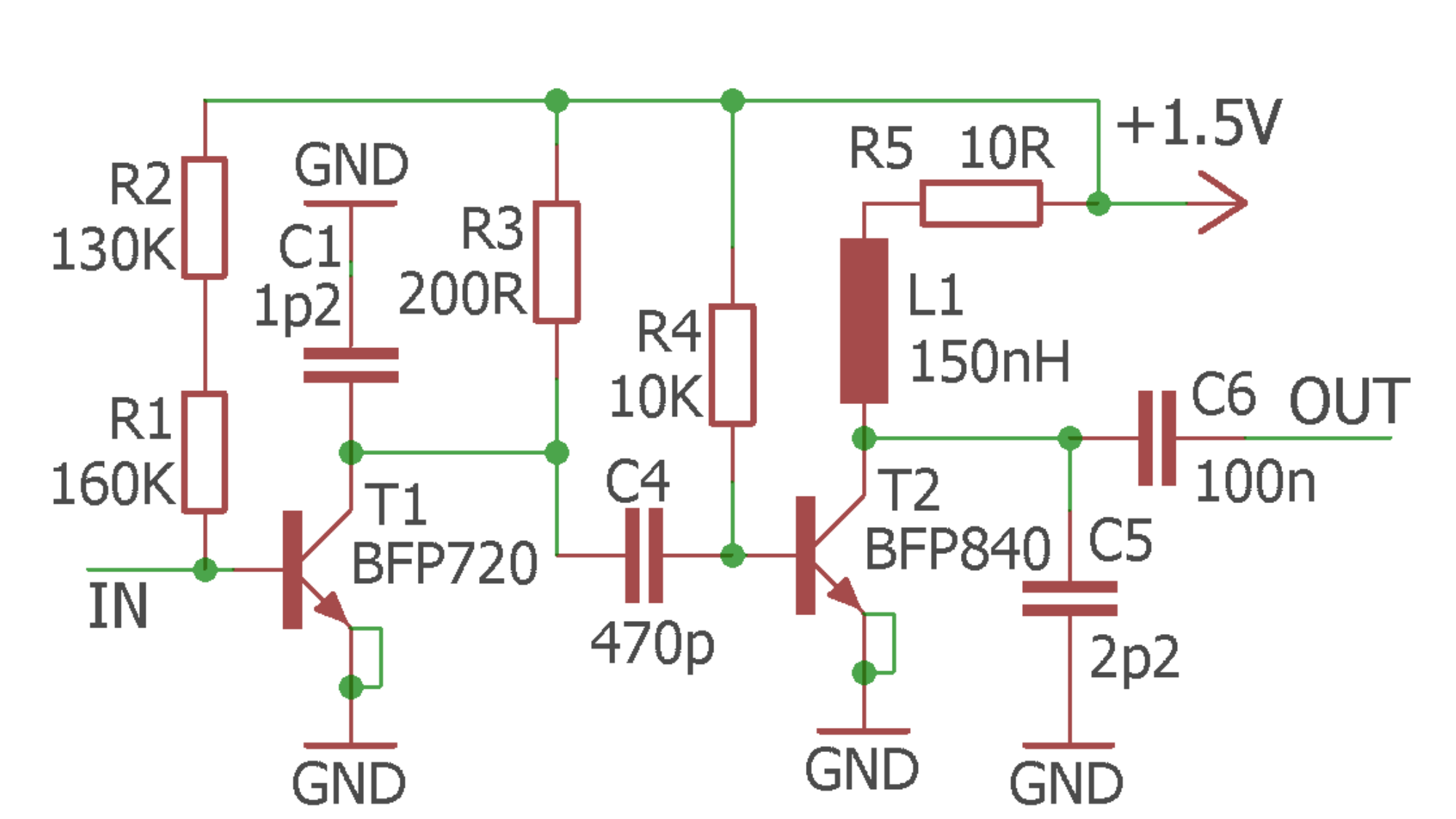}
\end{center}
\caption{Schematic of the front-end amplification system used in the test.}
\label{fig_fee}       
\end{figure}
\begin{figure}[h]
\begin{center}
  \includegraphics[width=0.45\textwidth]{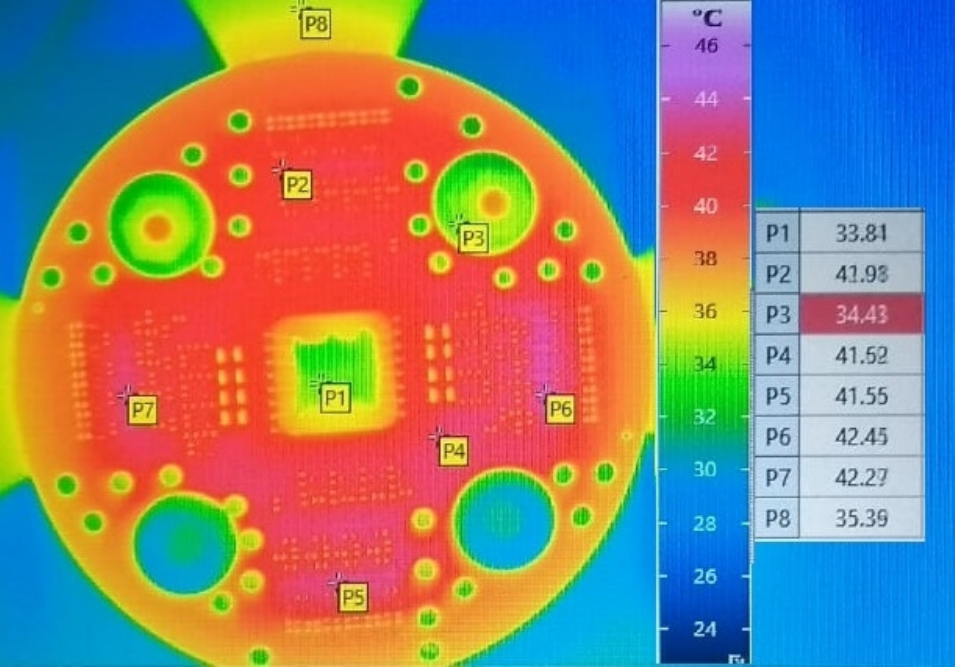}
\end{center}
\caption{A thermographic image of the PCB at operation voltage of 1.5 V equipped with an LGAD sensor. The transistor temperatures reach 42$^\circ$ Celsius ($e.g$ P5, P6) and are significantly higher than the sensor temperature (P1).}
\label{fig_temp_distribution}       
\end{figure}
\begin{figure}[h]
\begin{center}
  \includegraphics[width=0.45\textwidth]{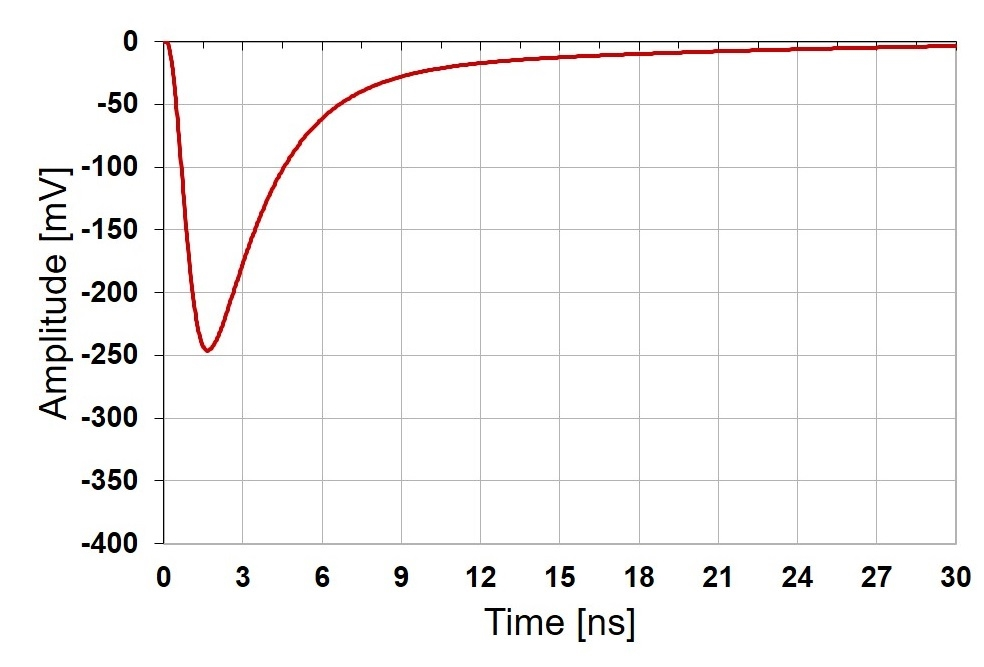}
\end{center}
\caption{Simulated response of the system shown in Fig.~\ref{fig_fee} to the corresponding to the LGAD detector signal response to a minimum ionising particle. Simulation parameters are described in the text.}
\label{fig_sim_signal}       
\end{figure}
The concept of two stages of amplification is shown in Fig.~\ref{fig_fee}.
It has been developed for the scCVD diamond sensors~\cite{Ref_Hades_di_mips} with a small strip capacitance of about 0.6~pF.
The first stage, mounted as close as possible to the sensor, with relatively large input impedance (kOhm), integrates the signal whereas the second stage provides L-R shaping determining the rise-time and baseline return.
The system is based on discrete transistor types BFP720 and BFP840.
A temperature stability test for the on board detector electronics has been carried out. Fig.~\ref{fig_temp_distribution} shows the temperature distribution on the circuit board in vacuum at nominal operation voltage of 1.5 V after 10 hours of operation.
While the transistor regions show a temperature of about 42$^\circ$ Celsius, the sensor remains cooler.
Tests carried out in a dark container in air at room temperature showed a similar temperature distribution, and the absolute values did not differ by more than 3 degrees.
Therefore, the beam tests were carried out without active cooling.
However, for the final T$_0$ detector a cooling system is planned.\\ 
Using the diagram presented in Fig.~\ref{fig_fee} the system's response to the LGAD signal for MIPs was simulated and is shown in Fig.~\ref{fig_sim_signal}.
As an input to the simulation, a fast current pulse with an amplitude of -16~$\mu$A and a fall time of 1.2~ns was used.
It corresponds to a charge of about 10~fC, which is generated by the LGAD sensor with a gain of 20 after passing of a minimal ionizing particle.
The transistor temperature used in this simulation was set to 40$^\circ$ Celsius.
For the mentioned input signal and the operational temperature, the simulated signal-to-noise (S/N) ratio of 163 and the noise level (RMS) of about 1.6~mV were obtained.
Laboratory measurements show a noise level of 1.7-1.8~mV (RMS).
The detector described above was exposed to a proton beam of 1.92~GeV kinetic energy at COSY in J\"{u}lich.
The analog signals from the LGAD sensor, operated at a gain of 20 after two amplification stages biased at 1.5~V, are illustrated in Fig.~\ref{fig_scope}.
As seen in the picture, the amplitude's most probable value (MPV) is located around 250 mV, consistent with the amplitude of the simulated electronics response to signals from the LGAD detector.
The long tail present in the signal comes from incorrect matching of the second stage of the amplifier to the strip capacitance and can be improved significantly.
In our prototype system, only 16 strips, due to the limited channels on the PCB (see Fig.~\ref{fig_ufsd_on_pcb_large}), located in the middle area of each sensor were bonded to amplifiers.
Analog signals from the sensor after two amplification stages, with a shape illustrated in Fig.~\ref{fig_scope}, were sent to the signal discrimination boards, as shown in Fig.~\ref{fig_3d_setup}.
\begin{figure}[t]
\begin{center}
  \includegraphics[width=0.48\textwidth]{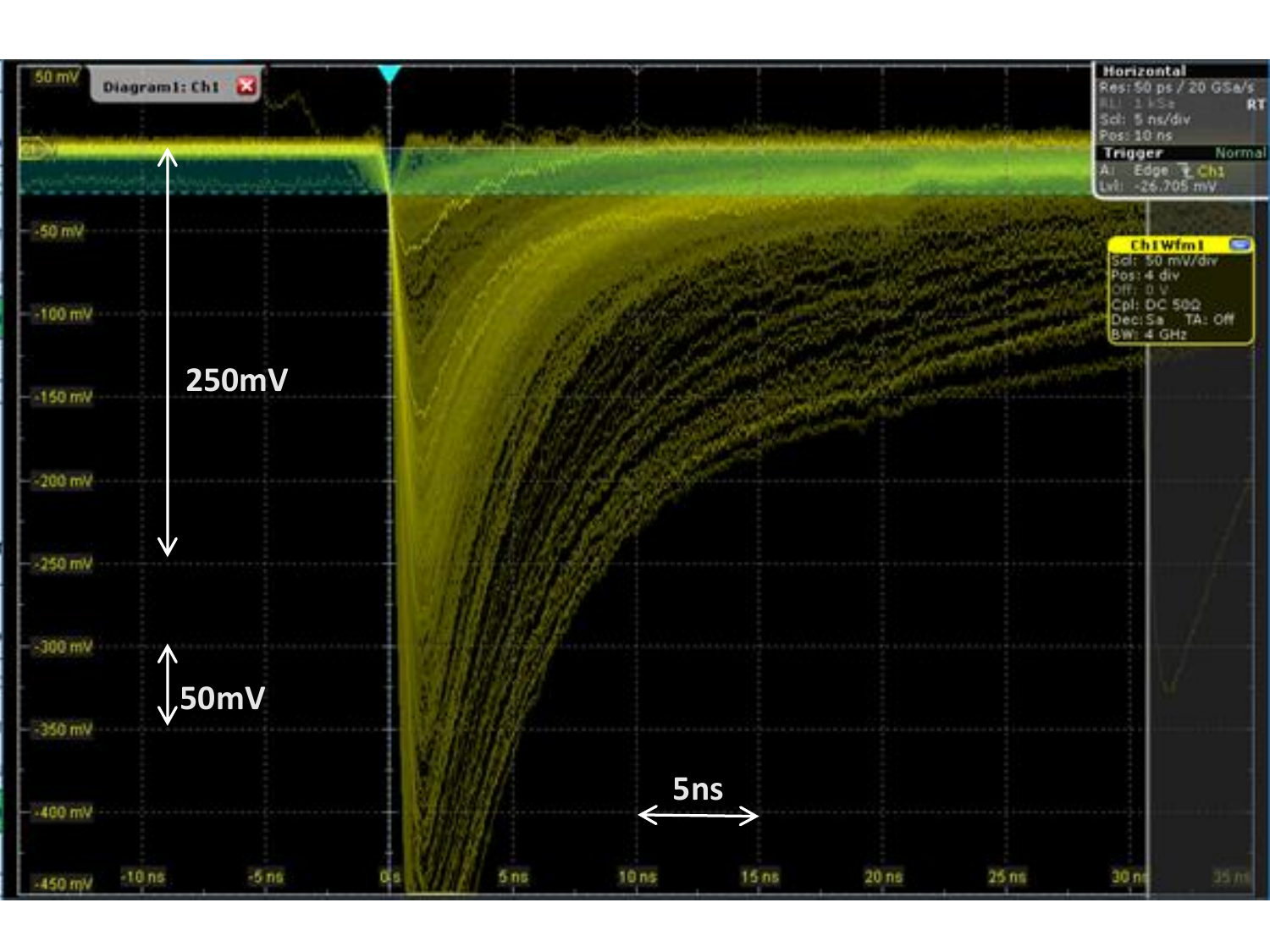}
\end{center}
\caption{Analog signals from the LGAD sensor after two stages of amplification obtained with proton beam of 1.92~GeV kinetic energy passing through the sensor. The shaping of the second amplifier was not correctly adjusted  to the sensor capacitance resulting in long signal tails. }
\label{fig_scope}       
\end{figure}
\begin{figure}[t]
\begin{center}
  \includegraphics[width=0.50\textwidth]{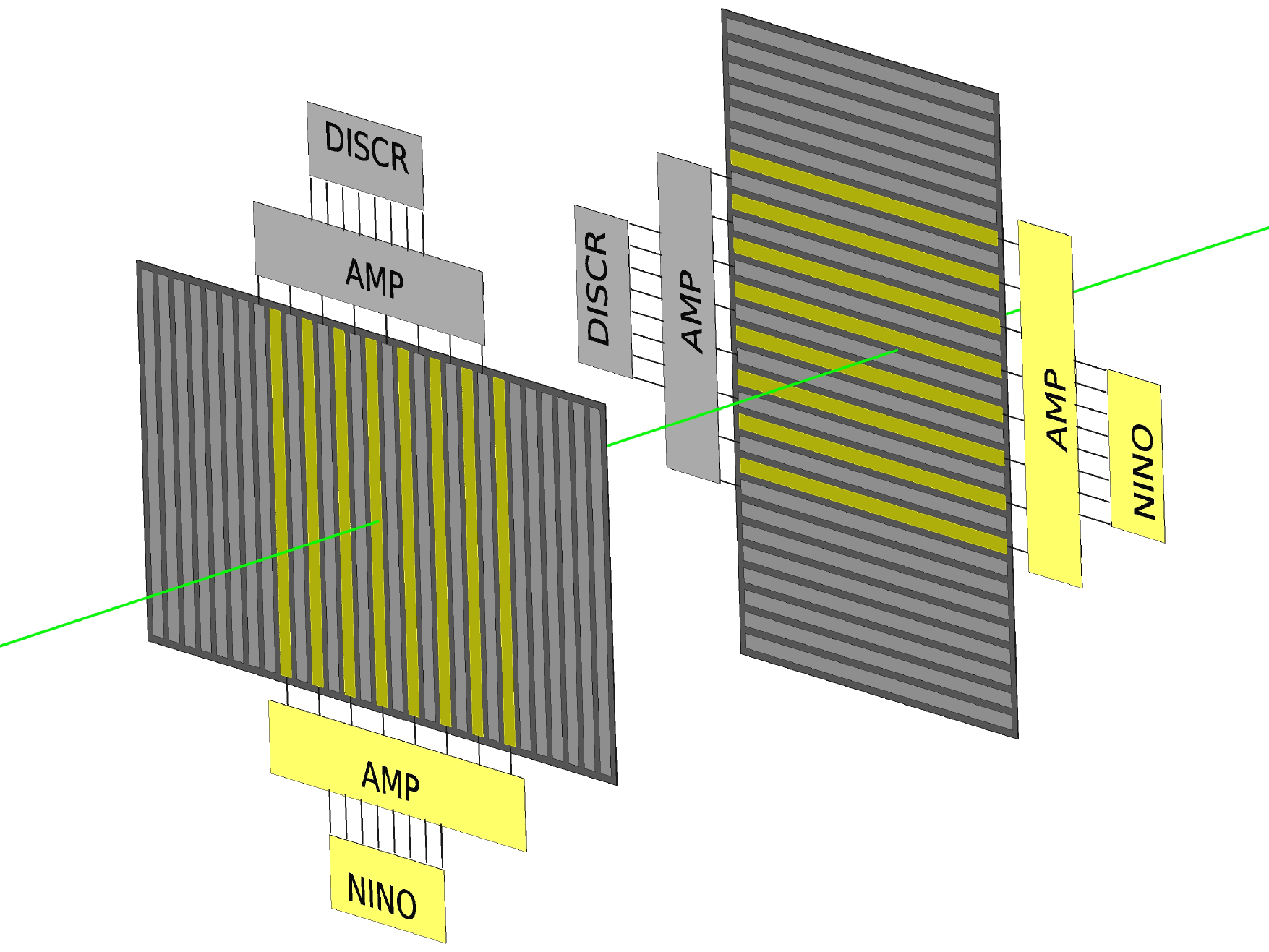}
\end{center}
\caption{Arrangement of two LGAD sensors forming a beam telescope used for the time precision determination. Strips were oriented orthogonal to provide two-dimensional position information. Two types of discriminator boards were connected to each sensor as explained in the text. Data from marked strips connected to the NINO discriminator boards (yellow-marked) was used for the analysis presented in this paper.}
\label{fig_3d_setup}       
\end{figure}
Two different discriminator boards were attached to each sensor, one based on the NINO chip~\cite{Ref_NINO} and the other utilizing discrete circuits. 
Only data from 8 channels of the NINO based boards are discussed in this paper while data from discrete discriminators were necessary to measure cluster size. 
The NINO based board offers eight input channels, and delivers eight fast Low Voltage Differential Signals~(LVDS). 
Each signal marks the leading edge discrimination time as well as the integrated signal coded into the pulse width. 
The NINO based discriminator boards were subsequently followed by a TDC system employing the FPGA-TDC concept~\cite{Ref_Trb3}.
To obtain two-dimensional position information, a setup consisting of two single sided strip sensors with orthogonal strip orientation was formed and aligned along the beam direction, as sketched in Fig.~\ref{fig_3d_setup}.
A photograph of this setup arranged as a beam telescope is shown in Fig.~\ref{fig_telescope}.
The beam was centered in the middle of the test setup with a beam spot diameter of about 1~cm and intensity ranging from 10$^5$ to 10$^6$~protons/s which corresponded to rates of 1-10~kHz per single strip, respectively.
\begin{figure}[th]
\begin{center}
  \includegraphics[width=0.48\textwidth]{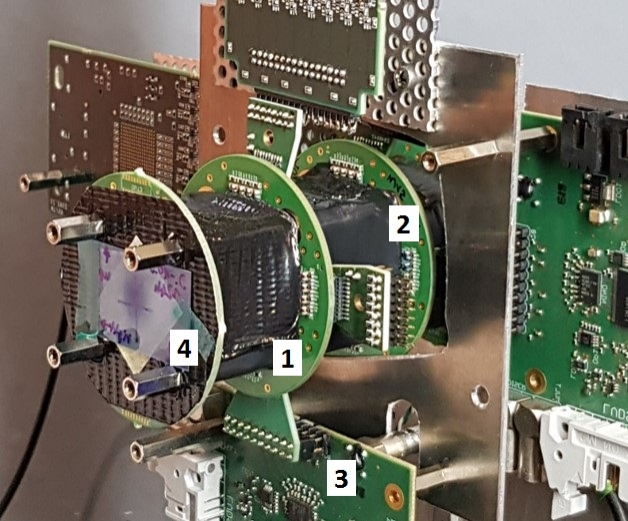}
\end{center}
\caption{Photograph of the telescope consisting of two aligned LGAD sensors mounted on PCBs and marked as (1) and (2), respectively, followed by discriminator boards~(3). Dummy PCBs~(4) were used in front and at the end of the telescope to provide light tightness. A dosimetry foil was mounted on the front PCB to visualize beam position and its alignment. In addition, the entire system was mounted on two connected linear stages, not shown in this photo, which made it possible to precisely position the system with respect to the beam.}
\label{fig_telescope}       
\end{figure}
The sensors were operated at atmospheric pressure and room temperature (without an active cooling).
The bias voltages applied to the W3 and W15 sensors were 300~V and 250~V, respectively.
The monitored leakage currents stayed relatively constant during the whole experimental period with values of 8~$\mu$A (W3 sensor) and 10~$\mu$A (W15 sensor).
A measurement of leakage current variations recorded on the W15 sensor within 14 days of operation in air without active cooling is shown in Fig.~\ref{fig_leakage_current}. \\
\begin{figure}[t]
\begin{center}
  \includegraphics[width=0.48\textwidth]{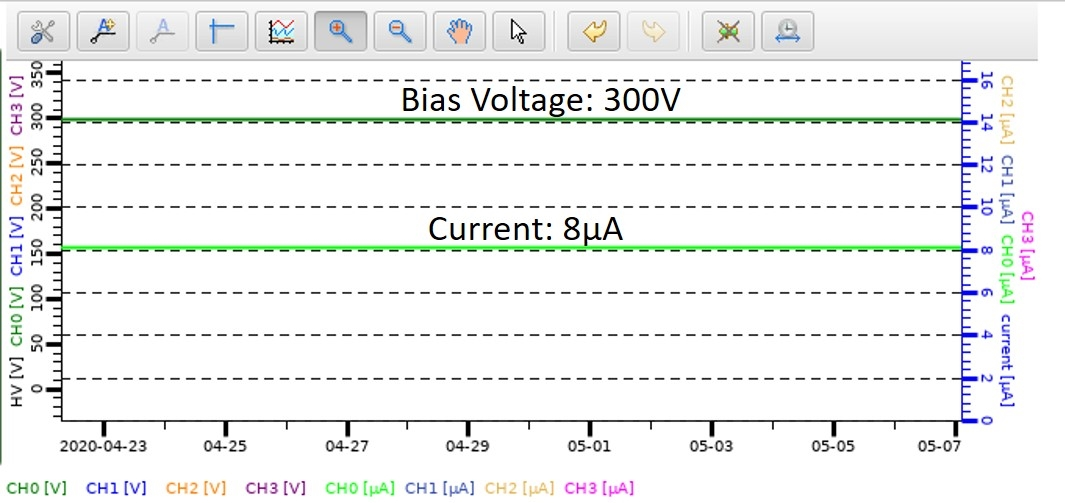}
\end{center}
\caption{Bias voltage/current stability trend recorded during 14 days on W15 sensor operated at 300V in air without active cooling.}
\label{fig_leakage_current}       
\end{figure}
\begin{figure}[t]
\begin{center}
\includegraphics[width=0.48\textwidth]{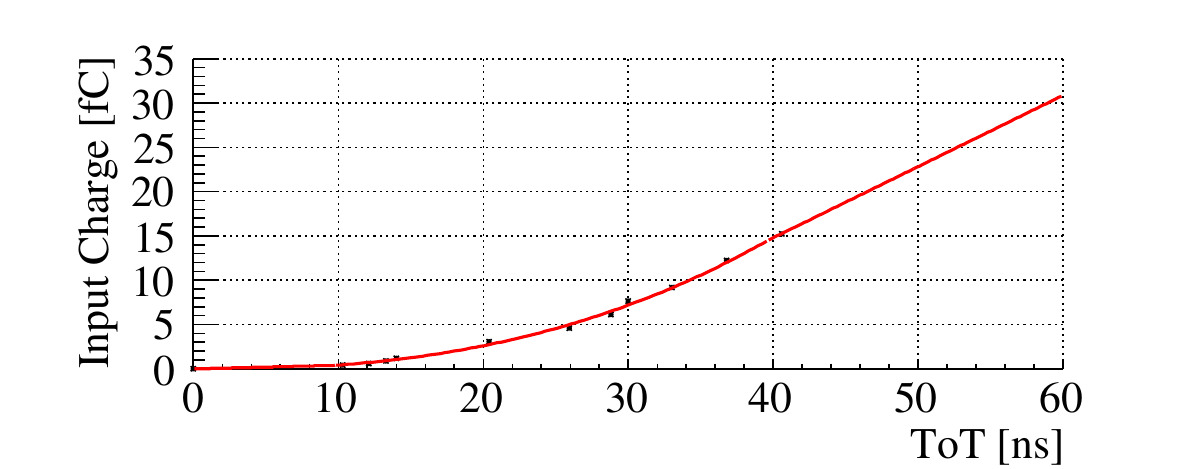}
\caption{Calibration curve describing the relation between the injected charge values and the measured ToT values. See text.}
\label{fig_charge_ToT_NINO}       
\end{center}
\end{figure}
As the amplification stages were installed very close to the sensor, the power dissipated by the FEE contributed to the sensor operation conditions. 
The sensors dissipated about 1.8~mW per sensor whereas the FEE contribution was much larger reaching 30~mW per readout channel.
Nevertheless,
the heat dissipated by both, the sensors and the FEE, did not influence significantly the performance of the prototype.

\section{Results}
\label{sec_results}
\subsection{Charge calibration and cluster size analysis}
\begin{figure}[h]
\begin{center}
  \includegraphics[width=0.48\textwidth]{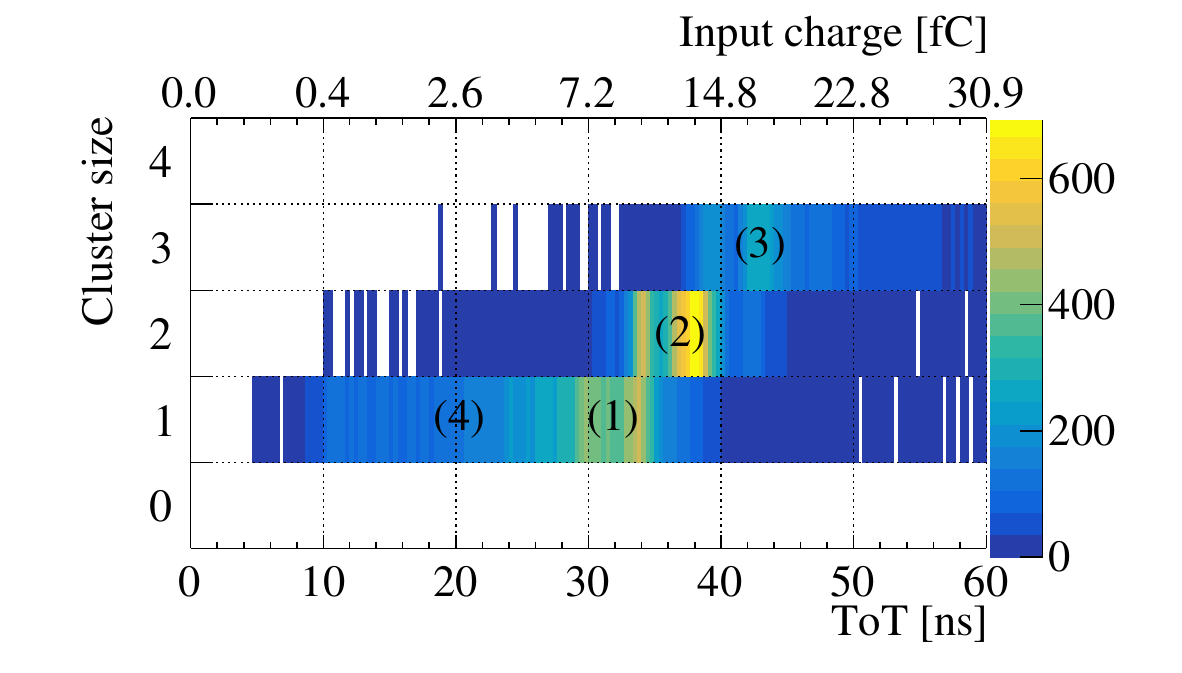}
\end{center}
\caption{Cross-talk representation caused by capacitive coupling occurring between adjacent detector strips. The main signal couples to neighboring strips and depending on their threshold levels can fire the discriminators of neighboring channels creating cluster size grater than 1, cases (2) and (3) or not, cases (1) and (4). The histogram was filled if the charge met the local maximum condition.}
\label{fig_mult_vs_ToT}       
\end{figure}

\begin{figure}[h]
\begin{center}
  \includegraphics[width=0.42\textwidth]{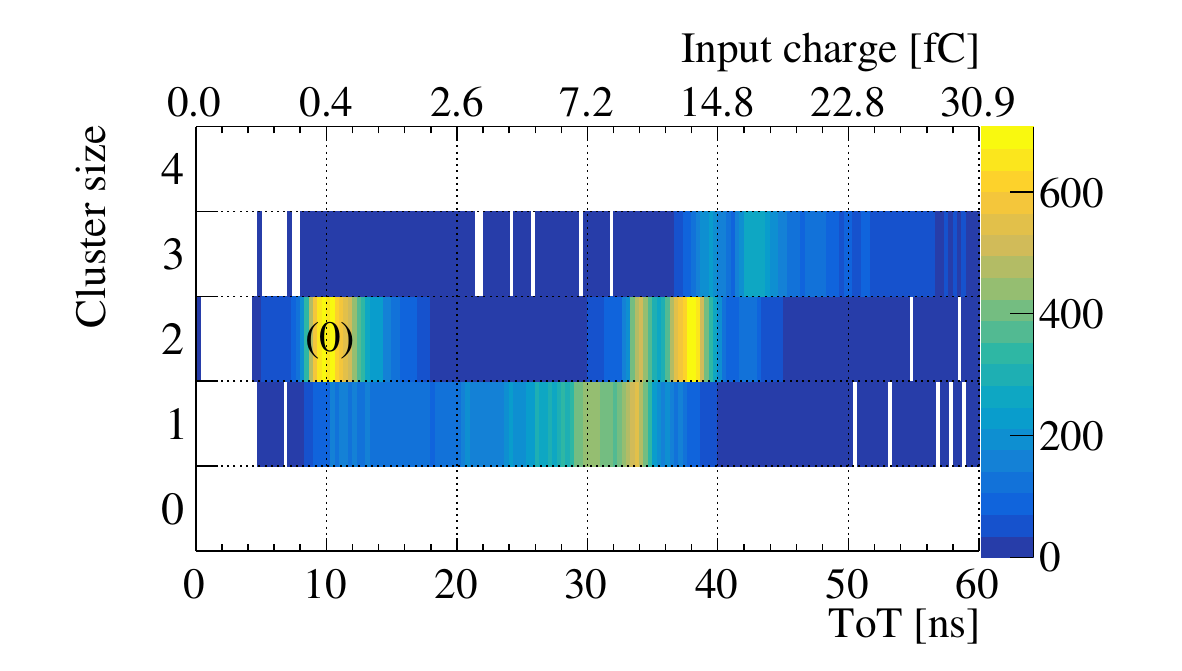}
\end{center}
\caption{The results of the analysis shown in Fig.~\ref{fig_mult_vs_ToT} without condition on local maximum. In this case, additional counts around low charge value of 0.4~fC appear, marked as (0), as a result of capacitive coupling when the main signal stemmed from one of the neighboring strips.}
\label{fig_mult_vs_ToT_wo_locmax}       
\end{figure}

Prior to the experiment, a Time over Threshold (ToT) calibration of the discriminator boards was performed.
Well defined signals with shapes compatible with a typical LGAD output, see Fig.~\ref{fig_scope}, were injected to discriminator inputs and the discriminator responses, so ToT values, were recorded.
The discriminator threshold levels used during this calibration procedure was 5~mV, identical to the ones used in experiment.
In addition, knowing the exact response of the electronics, as shown in Fig.~\ref{fig_sim_signal}, and knowing the amount of charge generated in the LGAD detector at a gain of 20, it was possible to obtain a relation between the injected charge and the ToT measured in the experiment.
A calibration curve obtained this way is shown in Fig.~\ref{fig_charge_ToT_NINO}.\\
In order to understand the distributions of measured charges by means of ToT, an analysis was carried out to examine the size of clusters.
The clustering procedure is based on measurements of a Time of Arrival (ToA) and charge from each individual strip.
It involves searching for the local maximum of charge in a given channel by comparing the charges of the three successive channels.
The result of this analysis for the selected single channel is shown in Fig~\ref{fig_mult_vs_ToT}.
It shows the dependence of the cluster size to the overall charge. 
This histogram was filled provided that the charge met the local maximum search condition.  
There are four regions as indicated in the figure.
Region (1) is characterized by a charge above 7~fC (ToT 29~ns) and a cluster size equal to 1.
In this case, we measure a single particle in the strip center and adjacent strips did not record signals above their thresholds.
Region (2) separates from region (1) when the charge begins to exceed 9~fC (ToT of 33~ns).
In this case, we observe a cluster size of 2 because a signal above threshold appears on one of the neighboring strips.
A cluster size of 3 appears in region (3) when the charge in the middle strip exceeds 12~fC (ToT 37~ns) and both adjacent strips register a signal above their thresholds.
The hit appears in region (4) when a particle passes in the area where the gain is lower than the nominal one.
We observe a decrease in the value of the measured charge up to a threshold value of 0.3~fC (ToT of 8~ns).
It can also be seen that our electronics is not efficient for signals generated in regions where the gain is 1 (no gain region).
Low charge values, around 0.4~fC, and cluster size 2 appear in the graph prepared without requesting the local maximum condition, see Fig.~\ref{fig_mult_vs_ToT_wo_locmax}.
This additional region around 0.4~fC, marked as (0), originates from capacitive coupling between adjacent strips, when the main charge was generated on one of the adjacent strips.
According to the capacitance estimations, also discussed in~\cite{Ref_Xray_test}, we expect about 5\% capacitive coupling to the adjacent strip, which agrees well with the observed distribution of cluster sizes.
The above described argumentation demonstrates that all charges below the value of 7~fC, ToT of 29~ns, originate either from capacitive coupling or from the area with reduced gain, and therefore are omitted when estimating the final precision of time measurement.
\subsection{Walk correction}
\label{sec_walk}
\begin{figure}[h]
\begin{center}
  \includegraphics[width=0.48\textwidth]{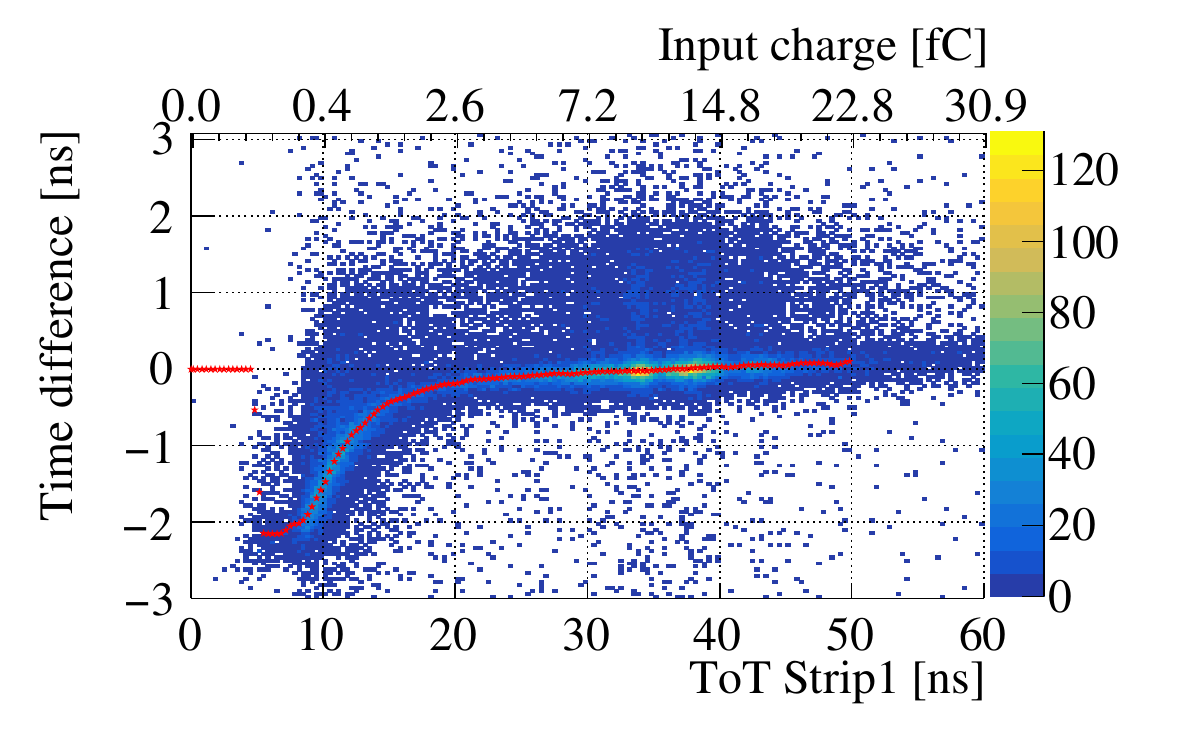}
  \includegraphics[width=0.48\textwidth]{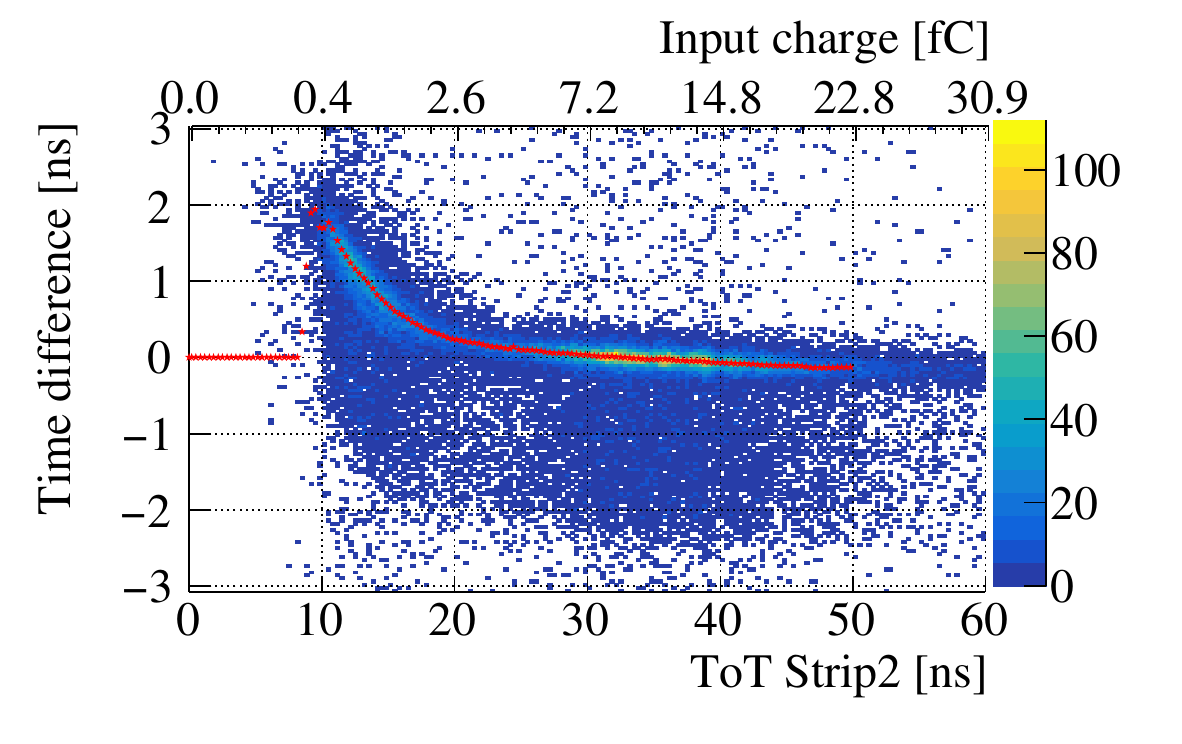}
\end{center}
\caption{Illustration of the walk effect: amplitude-dependent error in time difference of ToA measurements. The graphs represent the time differences between two readout channels from different LGAD sensors as a function of the ToT pulse width measured during the test experiment. Based on these graphs, the ToA correction parameters, represented by red dots, as a function of ToT were determined and used in the further steps of the analysis.}
\label{fig_walk_step1}       
\end{figure}

The determination of the time precision of the LGAD sensors was performed in several steps of the data analysis. In a first step ToA and ToT of protons passing through the telescope were measured and the difference of correlated ToA registered by the two sensors was calculated.
Fig.~\ref{fig_walk_step1} exhibits the typical dependence of difference of ToA from two strips from different sensors as a function of ToT values. 
As shown in Fig.~\ref{fig_scope}, the amplitude of the analog signals at the input to discriminators varies, thus time walk corrections based on the ToT measurement should be applied to improve the time precision.   
Fig.~\ref{fig_walk_step1} shows that, for signals with small amplitude, a significant walk effect reaching values up to 2~ns was observed. 

\begin{figure}[h]
\begin{center}
  \includegraphics[width=0.48\textwidth]{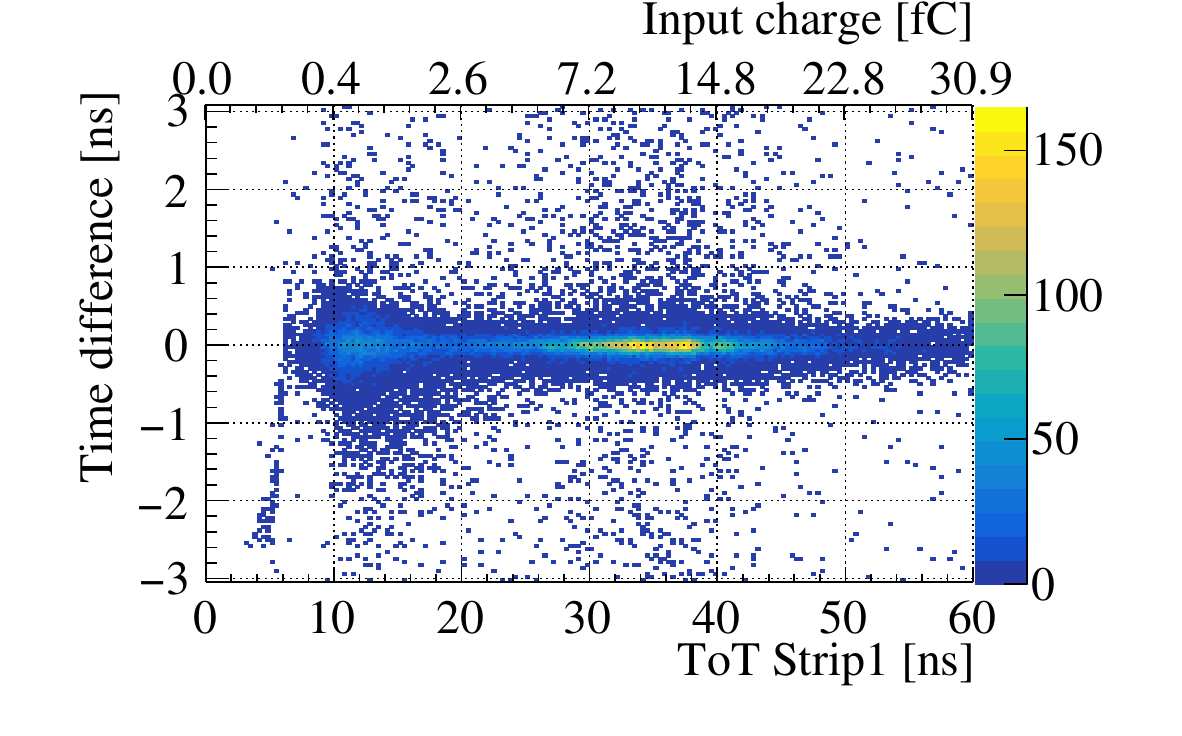}
  \includegraphics[width=0.48\textwidth]{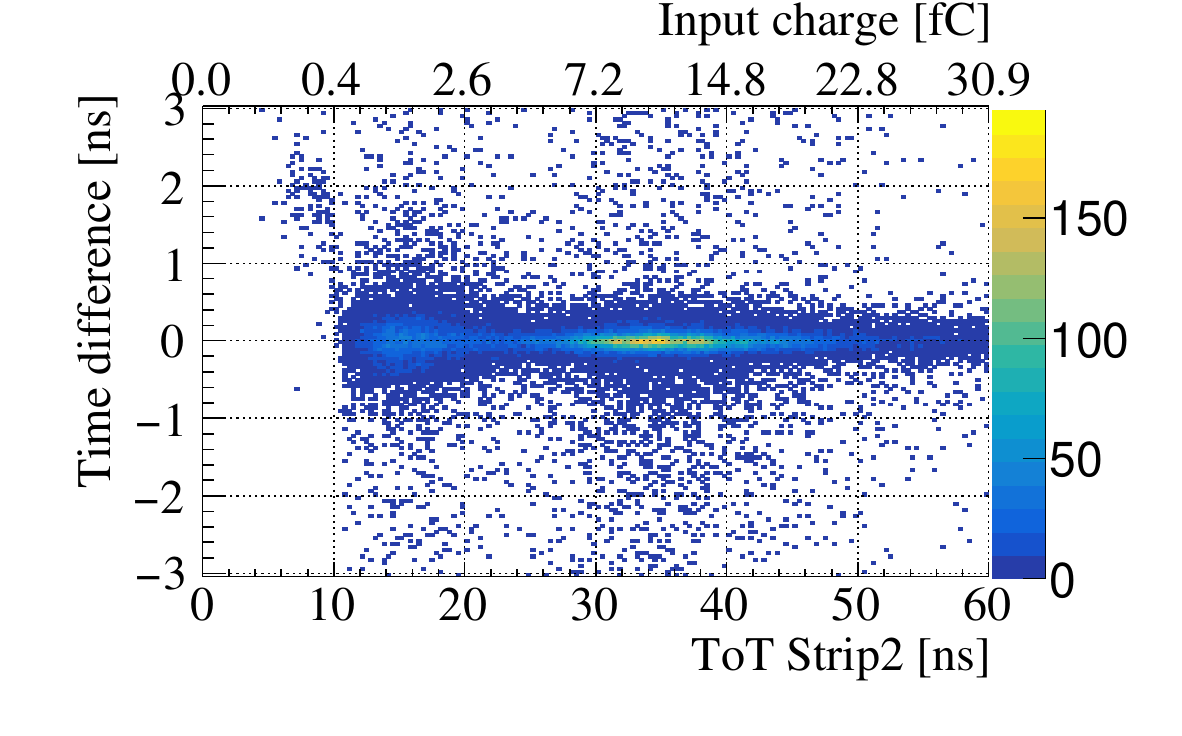}
\end{center}
\caption{The data presented in Fig.~\ref{fig_walk_step1}, after applying the time correction procedure (walk correction) as described in the text.}
\label{fig_walk_step2}       
\end{figure}

The walk correction parameters were deduced for each detector readout channel and the observed walk effect was compensated, in software, on event-by-event basis by measuring the ToT and correcting the ToA.
The result of this procedure is shown in Fig.~\ref{fig_walk_step2}.
The systematic dependence of the ToA difference on the ToT (input charge) could be entirely removed and hence allowed a precise time determination.

\subsection{Time precision}
\label{sec_Time_Precision}

\begin{figure}[t]
\begin{center}
  \includegraphics[width=0.48\textwidth]{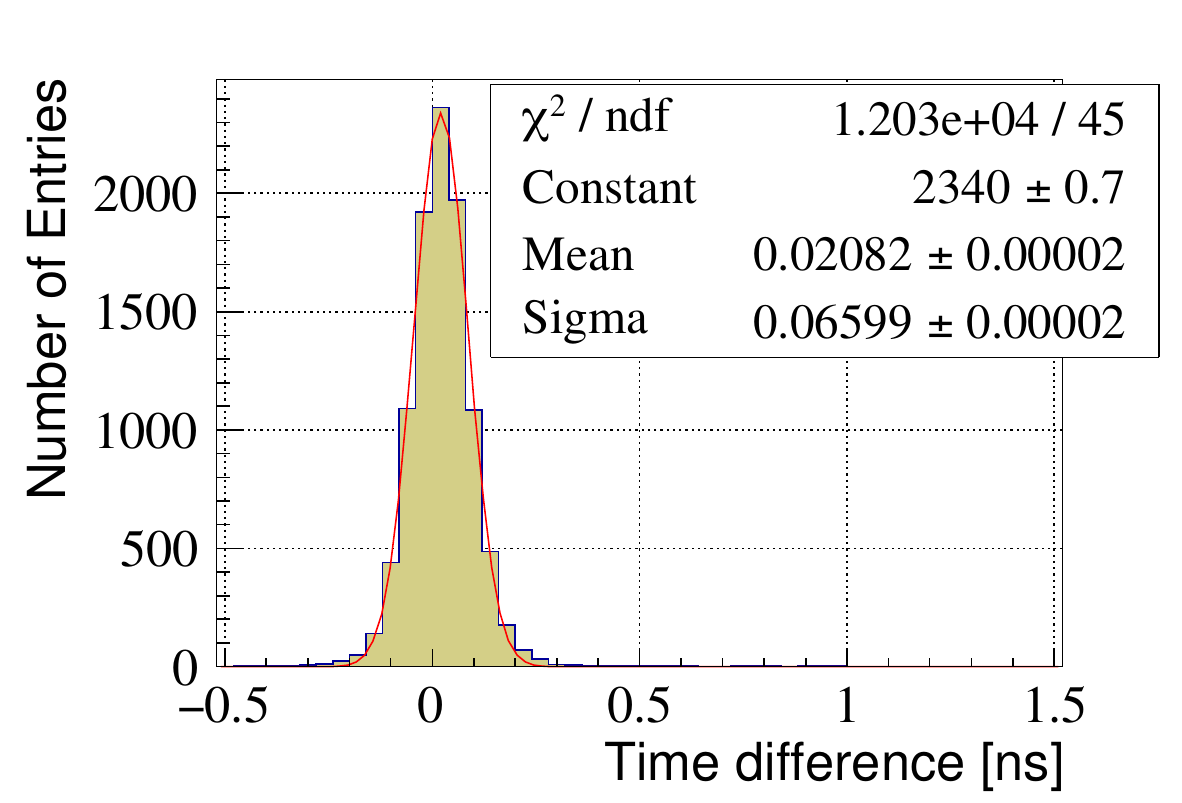}
\end{center}
\caption{Result of the time precision procedure based on a fit to the data presented in Fig.~\ref{fig_walk_step2}. As explained in the text, the lowest ToT values, which were outside of the range of interest, were excluded from the fit. The obtained time precision for this two-channel combination was 66~ps~/$\sqrt{2}$~=~47~ps.}
\label{fig_walk_step3}       
\end{figure}

With all results calibrated, the next step was to fit Gaussian function to 1D histograms of time difference between channels from different sensors.
In our example, after excluding low charge signals, the fits were performed by making projections onto the Y axis of the histograms, as shown in Fig.~\ref{fig_walk_step2}.
The outcome of this procedure for one pair of channels is shown in Fig.~\ref{fig_walk_step3}, resulting in a time precision of about 47~ps. 

\begin{figure}[t]
\begin{center}
  \includegraphics[width=0.48\textwidth]{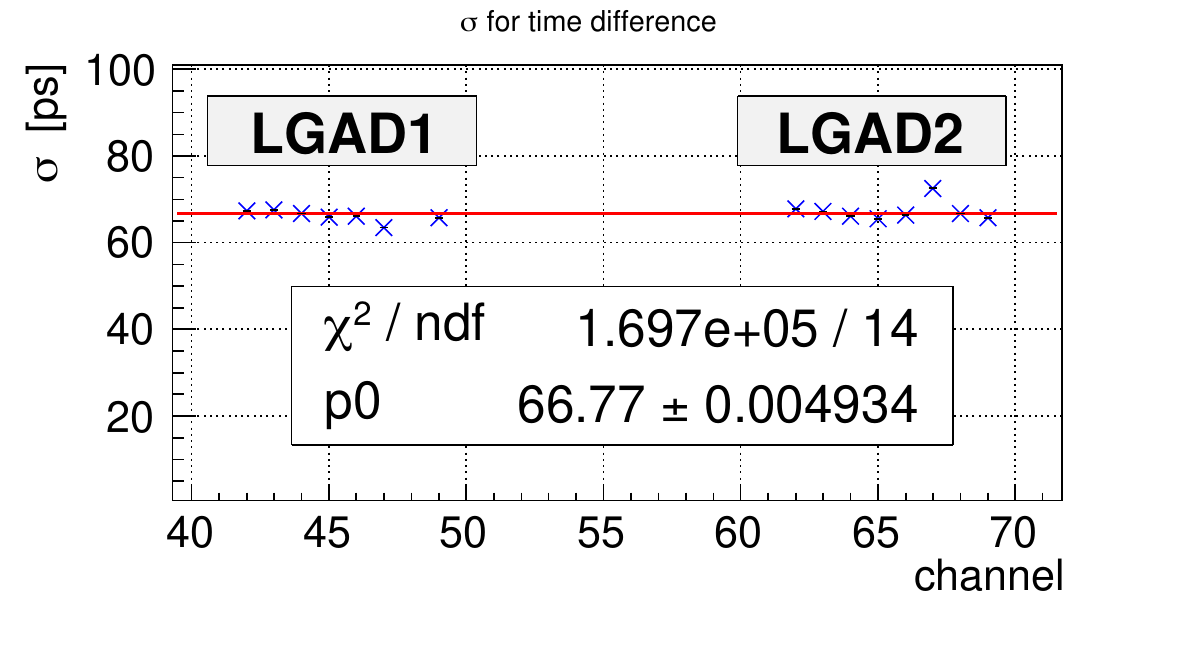}
\end{center}
\caption{A summary graph of time precision measurement showing results for 16 channels, 8 from each sensor under test. The averaged sigma value was 66.8~ps/$\sqrt{2}$=47.2~ps. Due to a malfunction one channel showed higher value and was excluded from the fit.}
\label{fig_Final_results}       
\end{figure}

The measured time precision for 16 channels, 8 from each LGAD sensor, are depicted in Fig.~\ref{fig_Final_results}. 
One channel, due to malfunction, provided a value far above the average.
For the other channels, the average value was 66.8~ps.
Given the fact that the result was obtained from the fit to the time difference between two channels, the time precision of a single channel is 66.8~ps/$\sqrt{2} =$ 47.2~ps, assuming similar precision of both channels. 

\section{Outlook}
\label{sec_conclusions}
We presented results obtained with a prototype T$_0$ telescope for the HADES experiment based on the single-sided LGAD strip sensors.
The setup included the readout system utilizing NINO leading-edge discriminators and a TDC system employing the FPGA-TDC concept. 
The time precision reaching 47~ps has been demonstrated, which is an excellent result for a system using leading edge technology.


The LGAD prototype sensors used in this test have a low fill factor of about 50\%.
Another drawback is the total thickness of a single sensor, which is 570~$\mu$m.
In order to improve the detection capabilities of the LGAD sensors and reduce the amount of material, a dedicated fabrication campaign is planned at FBK in 2020.
Sensors with different strip geometries with a no-gain-region of 30~$\mu$m will be manufactured.
In addition, it is foreseen to apply a thinning process to obtain sensors with a total thickness of about 200 $-$ 250~$\mu$m.
Such a thickness is essential to minimize the effect of multiple scattering of beam particles in the sensor itself. 
The T$_0$ detector mounted in front of the target in the HADES experiment will produce secondary particles and influence the beam focus.
Thus the total thickness of the active part of the two sensor layers, X and Y, should be below~500 $\mu$m to ensure proper focusing of the proton beam.
Although the tests were carried out without active cooling, for the final system we anticipate the use of cooling based on Peltier elements.
A specially designed PCB will ensure the transistor working temperature below 20 degrees and thus also lowering the temperature of the sensors itself.
The main PCB of the planned system hosts 94 amplifiers designed to work with two LGAD sensors with sizes up to 2.0~cm~x~2.2~cm. 
To realize larger size systems with sizes above several cm$^2$ an application-specific integrated circuit (ASIC) combined with LGADs will be of great importance and is planned in the near future.
%

\section{Acknowledgments}
\label{sec_acknowledgments}
We would like to gratefully thank the LGAD team at INFN-Torino, N. Cartiglia, V. Sola and M. Ferrero, and colleagues from Fondazione Bruno Kessler (FBK), M.~Centis Vignali, M. Boscardin, F. Ficorella, G. Paternoster, G Borghi, and O. H. Ali, and the  University of Trento G.F. Dalla Betta and L. Pancheri for providing us with sensors, help in detector preparation, strip sensor fabrication for the HADES experiment
and invaluable discussions.
The time precision results presented in this paper have been obtained at the COSY facility located in the Institute for Nuclear Physics at Reserach Center J{\"u}lich, Germany.
TU Darmstadt acknowledge support from the Deutsche Forschungsgemeinschaft (DFG, German Research Foundation) under grant No. GRK 2128.
The research presented here is a result of a R$\&$D project for the HADES experiment in the frame of FAIR Phase-0 supported by the GSI Helmholtzzentrum f{\"u}r Schwerionenforschung in Darmstadt, Germany.\\




\end{document}